\documentclass[review]{elsarticle}

\usepackage{lineno,hyperref}
\usepackage{commath}
\modulolinenumbers[5]

\journal{Neurocomputing}

\begin{document}

\begin{frontmatter}

\title{A Deep Neural Architecture for Harmonizing 3-D Input Data Analysis and Decision Making in Medical Imaging}

\author[addressDimitris]{Dimitrios Kollias}

\author[addressntua]{Anastasios Arsenos}
\author[addressntua]{Stefanos Kollias \corref{cor1}}

\cortext[cor1]{Corresponding author}
\ead{stefanos@cs.ntua.gr}

\address[addressDimitris]{School of Electronic Engineering \& Computer Science, Queen Mary University London, Mile End Rd, London, E14NS, United Kingdom}
\address[addressntua]{School of Electrical \& Computer Engineering, National Technical University Athens, Polytechnioupoli, Zografou, 15780, Greece}




\begin{abstract}
{Harmonizing the analysis of data, especially of 3-D image volumes, consisting of different number of slices and annotated per volume, is a significant problem in training and using  deep neural networks in various applications, including medical imaging. Moreover, unifying the decision making of the networks over different input datasets is crucial for the generation of rich data-driven knowledge and for trusted usage in the applications. This paper presents a new deep neural architecture, named RACNet, which includes routing and feature alignment steps and effectively handles different input lengths and single annotations of the 3-D image inputs, whilst providing highly accurate decisions. In addition,  through latent variable extraction from the trained RACNet, a set of anchors are generated providing further insight on the network's decision making. These can be used to enrich and unify data-driven knowledge extracted from different datasets. An extensive experimental study illustrates the above developments, focusing on COVID-19 diagnosis through analysis of 3-D chest CT scans from databases generated in different countries and medical centers.}
\end{abstract}


\end{frontmatter}

\nolinenumbers

\section{Introduction}\label{sec1}

In a variety of  applications,  input data are in the form of 3-D volumetric images, i.e., two dimensional image sequences which include different number of frames, or slices, and which are annotated in terms of a single label per sequence.  Such applications are, for example, 3-D chest CT scan analysis for pneumonia, COVID-19, or Lung cancer diagnosis \cite{ref1}, \cite{ref2}; 3-D magnetic resonance image (MRI) analysis for  Parkinson's, or Alzheimer's  disease prediction \cite{ref3}, \cite{ref34}; 3-D subject's movement analysis for action recognition \& Parkinson's detection \cite{ref5}; analysis of audiovisual data showing subject's behaviour for affect recognition \cite{ref6}; anomaly detection in nuclear power plants \cite{caliva}. Dealing with a single annotation per volumetric input and harmonizing the input variable length constitutes a significant problem when training Deep Neural Networks (DNNs) to perform the respective prediction, or classification task. 
 
Furthermore, in each of the above application fields, public, or private datasets are produced in different environments and contexts and are used to train deep learning systems to successfully perform the respective tasks. Extensive research is currently made on using data-driven  knowledge, extracted from a single, or from multiple datasets, so as to deal with other datasets. Transfer learning, domain adaptation, meta-learning, domain generalization, continual or life long learning are specific topics of this research, based on different conditions related to the considered datasets  \cite{ref10}. An additional condition can be that some, or all of the datasets may not be available during continual learning, due for example to privacy, or General Data Protection Regulation (GDPR) issues. In such cases it can be possible to perform diagnosis by only sharing some data-driven knowledge, like the  weights of independently trained  DNNs. 


COVID-19 diagnosis based on medical image analysis is the application domain examined in this paper.  
Various methods have been proposed to diagnose COVID-19, using analysis of chest x-rays, or CT scans. In particular, chest 3-D CT images can be used for precise COVID-19 early diagnosis.
Recent approaches target segmentation and automatic detection of the pneumonia region in lungs and  subsequent prediction of anomalies related to COVID-19 \cite{wang2021deep}. Common anomalies are multiple ground-glass opacity, consolidation, and interlobular septal thickening in both lungs, which are mostly distributed under the pleura. 

 Such approaches require large training datasets. A few databases with CT scans have been recently developed \cite{zhang2020clinically}, \cite{ref13}. However, a rather fragmented approach is followed: research is based on specific datasets, provided by small, or larger numbers of hospitals, with no proof of good performance generalization over different datasets and clinical environments. Moreover, many datasets are  small, in terms of total CT scans, or scan slices, or COVID-19 annotated CT scans,  or number of patients  \cite{yang2020covid}. In this paper we use a new very large database,  COV19-CT-DB, which we have developed, including chest 3-D CT scans, aggregated from different hospitals. In particular, it includes 7,750 3-D CT scans, annotated for COVID-19 infection; 1,650 are COVID-19 cases and 6,100 are non-COVID-19 cases. The 3-D CT scans consist of different numbers of CT slices, ranging from 50 to 700, totalling around 2,500,000 CT slices. Part of the database was successfully used in a recently held Competition  \cite{ref12}. The whole database is being currently made  available to the research community through our website. 
 
In the paper, we develop a deep neural architecture able to: i) analyze the 3-D CT scan inputs, ii) effectively handle the problem that each CT scan consists of a different number of CT slices and iii) provide a very high performance, when used on COV19-CT-DB and on other public datasets for COVID-19 diagnosis.
 RoutingAlignCovidNet (RACNet) is
  a CNN-RNN  architecture \cite{kollias2018deep} that is modified to include  routing and feature alignment steps that dynamically select the specific RNN outputs to be fed to the dense (fully connected) layers for decision making, i.e., COVID-19 diagnosis. 

In addition, we extract latent variables from the  trained RACNet and derive a set of anchors  which can provide insight into the network's data driven knowledge. Moreover, these anchors
 are used for unification with other datasets, thus developing a  continual learning framework which does not require  sharing of the input datasets.

  The rest of the paper is organized as follows. Related work is presented in Section II.  Section III provides a short description of the  COV19-CT-DB  database. The RACNet architecture is described in Section IV. Section V includes the experimental study presenting evaluation of the performance of RACNet when trained with COV19-CT-DB and then refined on other public databases. Conclusions and future work are presented in Section VI.

\section{Related Work}

A variety of 3-D CNN models have been used for detecting COVID-19 and distinguishing it from other common pneumonia (CP) and normal cases, using volumetric 3-D CT scans. In \cite{jaiswal2020classification}, a pretrained  DenseNet-201 was trained with CT scan images to classify them to the COVID-19, or non-COVID-19 category. The network's performance was compared to that of pre-trained and fine-tuned VGG16, ResNet152V2, and Inception-ResNetV2 networks. 
In \cite{khadidos2020analysis} a CNN plus RNN network was used, taking as input CT scan images and discriminating between COVID-19 and non-COVID-19 cases. 
In \cite{amyar2020multi}, the authors developed a multi-task architecture consisting of a (common) encoder that takes a 3-D CT scan as input and i) a decoder that reconstructs it; ii) a second decoder that performs COVID lesion segmentation; and iii) a multi-layer perceptron for classification between COVID and non-COVID categories.

In \cite{ref2}, a weakly supervised deep learning framework was presented using 3-D CT volumes for COVID-19 classification and lesion localization. A pre-trained UNet was utilized for segmenting the lung region of each CT scan slice; the latter was fed into a 3-D DNN that provided the classification outputs; the COVID-19 lesions were localized by combining the activation regions in the DNN and some connected components in unsupervised way. 
\cite{paperAAAI} first used 3D models, such as  ResNet3D101 and DenseNet3D121, to establish the baseline performance. Then it proposed a differentiable neural architecture search (DNAS) framework to automatically search the 3D DL models for CT scan classification. It has published the used  training, validation and test datasets for future research.

\section{The COV19-CT-DB Database}

COV19-CT-DB includes 3-D chest CT scans  annotated for existence of COVID-19. Data collection was conducted in the period from September 1 2020 to November 30 2021.  It consists of 1,650 COVID and 6,100 non-COVID chest CT scan series, which correspond to a high number of patients (more than 1150) and subjects (more than 2600). In total, 724,273 slices correspond to the CT scans of the COVID-19 category and 1,775,727 slices correspond to the non COVID-19 category. 

Annotation of each CT slice has been performed by 4 very experienced (each with over 20 years of experience) medical experts; two radiologists and two pulmonologists. Labels provided by the 4 experts showed a high degree of agreement (around 98\%). Each of the 3-D scans includes different number of slices, ranging from 50 to 700. This variation in number of slices
is due to context of CT scanning. The context is defined in terms of various factors, such as the accuracy asked by the doctor who ordered the scan, the characteristics of the CT scanner that is used, or specific subject’s features, e.g., weight and age.

\begin{figure*}[h!]
\centering
\includegraphics[height=0.19\linewidth]{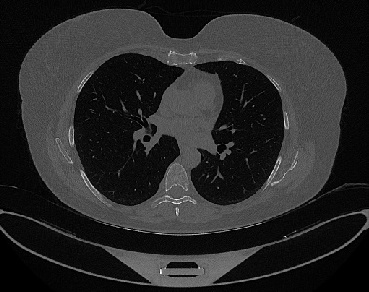}
\includegraphics[height=0.19\linewidth]{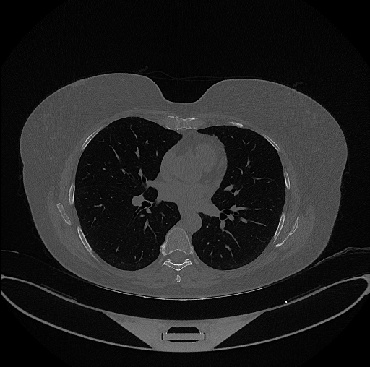}
\includegraphics[height=0.19\linewidth]{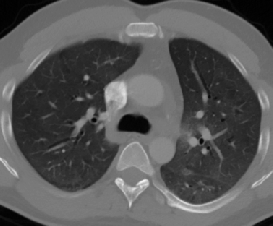}
\includegraphics[height=0.19\linewidth]{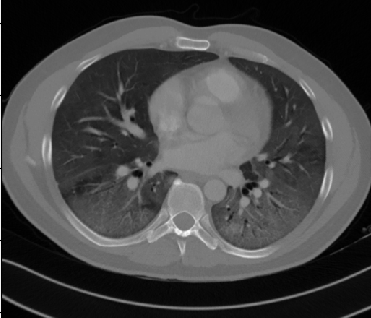}
\caption{Four CT scan slices, two from a non-COVID-19 CT scan, on the left and two from a COVID-19 scan, on the right, including bilateral ground glass regions in lower lung lobes.} 
\label{samples}
\end{figure*}

\begin{figure}[h!]
\centering
\includegraphics[height=3.9cm]{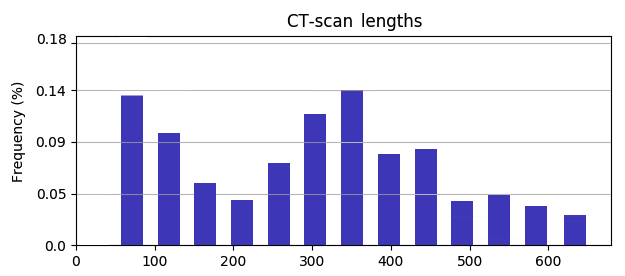}
\caption{Histogram of CT scan lengths}
\label{histograms}
\end{figure}

Figure \ref{samples} shows four CT scan slices, two from a non-COVID-19 CT scan, on the left and two from a COVID-19 scan, on the right. Bilateral ground glass regions are seen especially in lower lung lobes in the COVID-19 slices.

\section{RACNet: The Proposed Architecture}

\subsection{3-D Analysis and COVID-19 Diagnosis}

Let us focus on the COVID-19 diagnosis problem, so that we specifically define the input data characteristics. The input sequence is a 3-D signal, consisting of a series of chest CT slices, i.e., 2-D images, the number of which is varying. 
The 3-D signal can be handled using a 3-D CNN architecture, such as a 3-D ResNet. However, handling the different input lengths, i.e., the different number of slices that each CT scan contains, can only be tackled in some ad-hoc way, by selecting a fixed input length and removing slices when a larger length is met, or duplicating slices when the input contains a smaller number of slices. 
The 3-D signal could alternatively be handled using different Multiple Instance Learning methods \cite{carbonneau2018multiple}. Nevertheless, this does not fit this case, as the problem is not to identify one or more CT slices that illustrate COVID-19 occurrence; it is to learn  doctors' diagnosis making,  after examining the whole 3-D CT scan.

\begin{figure*}[h!]
\centering
\includegraphics[width=1.\linewidth]{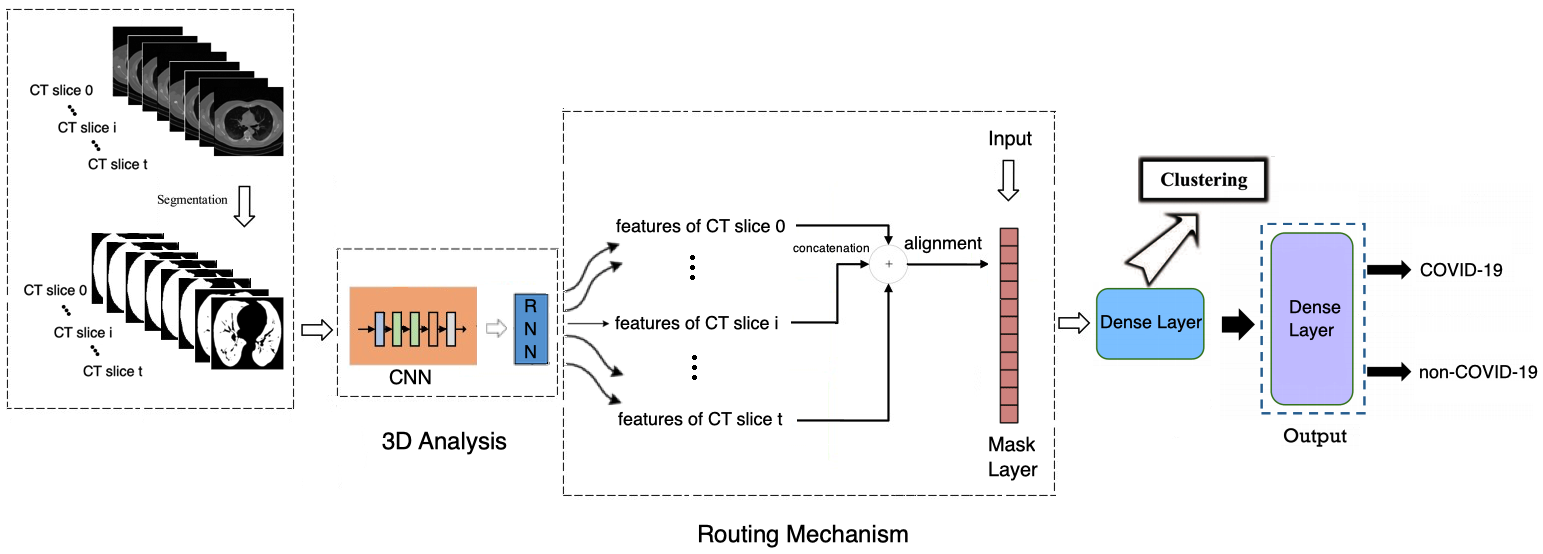}
\caption{The proposed Pipeline: A 3-D input composed of, up to $t$ chest CT slices is processed for COVID-19 diagnosis; 3-D analysis is performed by a CNN-RNN architecture, while a routing mechanism including an 'alignment' step and a Mask Layer handles the varying input length $t$. 
A dense layer follows, preceding the output
layer that provides the COVID-19 diagnosis; the neuron outputs of the dense layer are further analyzed through clustering to derive  a latent variable model and a related set of anchors  that provide further insight into the achieved decision making.} 
\label{methodology_iccv}
\end{figure*}

In the following we propose a CNN-RNN architecture, RoutingAlignClusterNet (RACNet), instead of a 3-D CNN one. By including a Mask Layer after the RNN part, RACNet dynamically selects  RNN outputs taking into account the input length, i.e., the number of slices of the analyzed CT scan. This is depicted in Figure \ref{methodology_iccv}, where $t$ denotes the maximum number of slices that appear among all available chest CT scans; the Mask Layer performs a dynamic routing procedure, as is explained below.

There are two modes of feeding input data to our model. In the first, segmentation of each 2-D slice is performed so as to detect the lung regions; then the resulting segmented image constitutes the input to the CNN. In the second, the whole unsegmented 2-D slices are fed as input to the CNN part. Both modes are studied in the presented experimental study of the paper.

At first all input CT scans are padded to have length $t$ (i.e., consist of $t$ slices). Then the CNN part performs local, per 2-D slice, analysis, extracting features mainly from the lung regions. The target is to make diagnosis using the whole 3-D CT scan series, similarly to the way medical experts provide the  annotation. The RNN part provides this decision, analyzing the CNN features of the whole 3-D CT scan, sequentially moving from slice $0$ to slice $t$. 

As shown in Figure \ref{methodology_iccv}, we get RNN features corresponding to each CT slice, from $0$ to $t$. We then concatenate these features and feed them to the Mask Layer. The original (before padding) length $l$ of the input series is transferred from the input to the Mask Layer to inform the routing process. During RACNet training, the routing mechanism performs dynamic selection of the RNN outputs/features. In particular, it selects as many of them as denoted by the length $l$ of the input series, keeping their values, while zeroing the values of the rest RNN outputs. In this way, it is routing only the
selected ones into the following dense layer. 

This can be done: a) by selecting the  first $l$ RNN outputs, or, b) through an 'alignment' step, i.e.,  by first placing the $l$ RNN outputs in equidistant positions in  $[0, t]$ and by then placing the remaining outputs in the in-between positions; the Mask gets their positions and performs routing of the respective RNN outputs to the following dense layer.

This dense layer learns to extract high level information from the concatenated RNN outputs. During training, we update only the weights that connect the dense layer neurons with the RNN outputs routed in the concatenated vector by the Mask layer. The remaining weights are updated whenever (i.e., in another input CT scan) respective RNN outputs are selected in the concatenated  vector by the Mask Layer. Loss function minimization is performed, as in networks with dynamic routing, by keeping the weights that do not participate in the routing process constant, and ignoring links that correspond to non-routed RNN outputs.       

 The final output layer follows, using a softmax activation function and providing the final COVID-19 diagnosis.


\subsection{The 'alignment' step}

Let us, for example, assume that a maximum input length of 700 CT scan slices is considered. For a specific input CT scan consisting of 50 slices, 650 duplicate slices are inserted so that the scan is made to contain 700 slices in total. During training, all 700 slices are fed to the CNN-RNN network. 

In the case where no 'alignment' is performed, the network's output is fed to the Mask layer which: i) zeroes the features corresponding to the 650 duplicate slices (slices 50-699), ii) lets the first 50 features (corresponding to original slices 0-49) keep their values. 


In the case where 'alignment' is performed, the features extracted from the CNN-RNN part are re-positioned as follows. The features corresponding to the 50 original slices (0-49) are placed in equidistant positions in $[0, 699]$. The rest features corresponding to the 650 duplicate slices are placed in the in-between positions. 
The operation of the Mask Layer is the same as when no 'alignment' is performed; it i) zeroes the features corresponding to the 650 duplicate slices (slices 50-699), ii) lets the other 50 features (corresponding to original slices 0-49) keep their values. 


In both cases, the 'masked features of CT slices 0-49' and the 'masked features of duplicate CT slices 50-699' are fed to the dense layer that precedes the output layer.

\subsection{Latent Variable Analysis and Anchor Set Generation} \label{clustersection}

In the proposed methodology we extract and further analyze, through clustering, the, say $L$, neuron outputs of the dense layer of the trained RACNet.  These latent variables carry high level, semantic information, which is used to generate the final classification at the output layer. We choose to discard the output layer and perform unsupervised analysis of these variables, so as to generate a representation that can provide further insight into the achieved decision making ability.

Let us assume that we feed the presented  architecture with a training dataset. For each  3-D CT scan input $k$, we extract the $L$ neuron outputs of the dense layer, forming a vector ${\textbf{v}}(k)$. In total, we get:

\begin{equation}
\label{eq:traininglatent}
\mathcal{V} = \big\{({\textbf{v}}(k), \ k=1,\ldots,N\big\} 
\end{equation}
where $N$ is the number of available training data. 

We aim to generate a concise representation of the $\textbf v$ vectors, that can be used as a backward model, to trace the most representative CT scan inputs for the performed diagnosis. This is achieved using a clustering algorithm, e.g.,  k-means++ \cite {arthur2006k}, which generates, 
say, $M$ clusters  ${Q} =\{\textbf{q}_1,\ldots,\textbf{q}_M\}$ by minimizing the following criterion: 
\begin{equation}
\label{eq:kmeans}
\widehat{{Q}}_{k\text{-means}} = \underset{{Q}}{\operatorname{arg\ min}} 
\sum_{i=1}^{M} \sum_{\mathbf{v}\in {V}}^{} 
{\norm{\textbf{v}-\textbf{$\mu$}_{i}}}^{2}
\end{equation}
where $\textbf{$\mu$}_{i}$ is the mean of $\textbf v$ values included in cluster $i$. 

Then, we compute each cluster center $\textbf{c}(i)$, generating the set $C$, which constitutes the targeted concise representation:

\begin{equation}
\label{eq: cluster centroid set}
\mathcal{C} = \big\{(\textbf{c}(i), \ i=1,\ldots,M\big\} 
\end{equation}

The CT scan inputs corresponding to the cluster centers can be then examined by medical experts, who can add  semantic information related to each one of them. 

The generated $C$ set can be used as an anchor set model assisting COVID-19  diagnosis in new subject cases. By testing the trained RACNet model on a new input CT scan case, we  will extract the corresponding  
$\textbf{v}$ vector of latent variables and will compute the euclidean distance between this vector and each cluster center, i.e., anchor, in $C$. As a consequence, the new input case is linked to the cluster center with the minimum euclidean distance and is annotated with the label of this center.  

The presented latent variable extraction and anchor generation can, therefore, be used to assist COVID-19 diagnosis in a rather efficient way; by computing  $M$ distances between $L$-dimensional vectors and selecting their minimum value. Moreover, by computing the respective cluster radii, we can provide confidence levels in addition to the confidence levels provided by the RACNet output layer. 

An additional advantage of this approach, when added to the main RACNet architecture, is that the latter needs retraining, or fine-tuning with new datasets, whenever such datasets become available, e.g., are generated by another hospital, or in another country. Due to privacy purposes, it is highly probable that it is not possible for different medical centers to share their datasets for retraining with all of them.
It can, however, be possible for different medical centers to share, or find on github, the best performing networks trained on others' data, as well as the respective anchor derived information. By continual aggregation of older and newer anchor sets, together with the respective trained (RACNet) networks, they can generate common enriched data-driven representations.

Regarding implementation of the proposed methodology: i) we used EfficientNetB0 as CNN model, stacking a global average pooling layer on top, a batch normalization layer and dropout (with keep probability 0.8); ii) we used a single one-directional GRU consisting of 128 units as RNN model; iii) the first dense layer consisted of 128 hidden units. Our model was fed with 3-D CT scans composed of CT slices; each slice was resized from its original size of $512 \times 512$ to  $256 \times 256$. 

In k-means clustering, we  tried values of $k \in \{2,..,25\}$, whilst evaluating the performance of our model on the validation set of COV19-CT-DB; k = 11 (7 clusters labelled as COVID-19 and 4 as non COVID-19) provided the optimal performance. 
Batch size was equal to 5 (i.e, at each iteration our model processed 5 CT scans) and the input length 't' (see Figure \ref{methodology_iccv}) was 700 (the maximum number of slices found across all CT scans). Loss function was the softmax cross entropy. Adam optimizer was used with learning rate $10^{-4}$. Training was performed on a Tesla V100 32GB GPU.

\section{Experimental Study}

This section describes a set of experiments evaluating the performance  of the proposed approach. 

At  first, we compare the performance of RACNet with the performance of  other types of networks, i.e., 3-D CNNs  and CNN-RNNs, using non-segmented CT scans of the COV19-CT-DB database; we show that it outperforms these networks in COVID-19 diagnosis. 
Furthermore, we perform ablation studies that illustrate the contribution of the various components of RACNet. In particular, we focus on: the routing mechanism ('alignment' step and mask); the choice of CNN model; the number of units in the dense layer; the use of 3-D convolution layers instead of RNN.

Next, we segment all CT scans in COV19-CT-DB and use the segmented CT scans for training RACNet. In particular, in order to compare RACNet's performance to the performance of the three winning methods in a ICCV 2021 Competition on COVID-19 diagnosis  \cite{ref12}, we adopted the same part of the COV19-CT-DB, composed of 5000 CT scans which was used in this Competition. The dataset was split in training, validation and testing sets. The training set contained, in total, 1552 3-D CT scans corresponding to 707 COVID-19 cases and 845 non-COVID-19 cases. The validation set consisted of 374 3-D CT scans, 165 of which represented COVID-19 cases and 209 represented non-COVID-19 cases. Finally the test set included 443 COVID-19 and 3012 non COVID-19 CT scans.
We make a comparison of RACNet’s performance, trained with the above dataset, to the performance of the three winning methods in the above-mentioned Competition, showing that RACNet outperforms all three methods. 

Then we evaluate RACNet's performance on another publicly available CT-scan database,  CC-CCII, and show the improved performance when compared to that of the recently published state-of-the-art DNAS framework referenced in Section II \cite{paperAAAI}.

Finally, we perform latent variable extraction from the trained RACNet, deriving the set of anchors and subsequently use them (in addition to RACNet) to derive a unified model over the COV19-CT-DB and the CC-CCII   \cite{ref13}.

\subsection{Comparison with 3-D CNN and CNN-RNN}

At first, we considered a 3-D CNN, the 3-D ResNet-50. According to \cite{hara3dcnns} 3-D CNNs require a large number of labeled image sequences to learn the 3-D kernels. Therefore, we used a pre-trained for action recognition 3-D ResNet-50 \cite{han2020accurate} and further trained it on COV19-CT-DB. 
We also  considered the MedicalNet\cite{chen2019med3d}, a 3-D ResNet-34 network trained for pulmonary nodule classification. MedicalNet has been trained with CT scans (i.e., the same input considered in this paper). We used it as a pre-trained network and further trained it on COV19-CT-DB. 

Finally we utilized a conventional CNN-RNN that outputs a probability for each CT scan slice and is followed by a voting scheme that makes the final decision; the voting scheme is either a majority voting, or an at-least one voting (i.e., if at least one slice in the scan is predicted as COVID-19, then the whole CT scan is diagnosed with COVID-19, otherwise it is diagnosed with non-COVID-19); the CNN and RNN parts of this network were the same as the respective parts of COV19-RACNet, i.e., EfficientNetB0 and single-directional GRU (we also considered a dense layer between the RNN and the output layer, but the achieved performance was worse). 

Table \ref{3dcnn_rnn} compares the performance of these networks to the performance of the proposed RACNet over the non-segmented COV19-CT-DB database. It  can be seen  that RACNet outperforms all of them, in terms of both accuracy and F1 score, for both COVID-19 and non COVID-19 categories. Particularly,  RACNet -although a lighter model- outperforms the 3-D ResNet-50 that has been pre-trained on a large action database. It also outperforms MedicalNet that has been pre-trained, not on action recognition (which is another task), but on a task similar to this paper's one. This shows that the proposed model structure, although  lighter than the 3-D ones (around 7M parameters versus about 63M and 46M respectively) and although it is not  pre-trained on another task, achieves a better performance. 

In addition, RACNet outperforms the CNN-RNN network; the main downside of that model is that there exists only one label for the whole CT scan and there are no labels for each CT scan slice. It should be mentioned that, in contrast to this, RACNet analyzes the whole CT scan, based on information extracted from each slice, as was described in the former Section.

\begin{table}[t]
\caption{Performance comparison between RACNet and other state-of-the-art structures on COV19-CT-DB database (non-segmented data)}
\label{3dcnn_rnn}
\centering
\scalebox{1.}{
\begin{tabular}{|c|c|c|c|}
\hline
Method   & \multicolumn{2}{c|}{Accuracy} & \multicolumn{1}{c|}{F1 Score}\\
\hline
 & COVID & non COVID &  \\ 
 \hline
 \hline
3D ResNet-50 \cite{han2020accurate}   & 0.74  & 0.80    &  0.82 \\
\hline
MedicalNet \cite{chen2019med3d}   & 0.78 & 0.83  &  0.86 \\
\hline
EfficientNet-GRU   & 0.73 & 0.80  &  0.82  \\
\hline
\hline
RACNet   & \textbf{0.82} &  \textbf{0.86} &  \textbf{0.90}   \\
\hline
\end{tabular}
}
\end{table}

\subsection{Ablation Studies} 
i) We studied the effect of the routing mechanism on our developed model's performance. We compared the performance of RACNet when the mask was used (and thus, only the RNN outputs corresponding to true input slices were routed to the dense and output layers) and when the mask was not used (and thus, inserted duplicates of input slices were also fed to the dense layers). Table \ref{ablations} illustrates the efficacy of the proposed mask, compared to the case of no masking. The mask effectively filtered out   unnecessary and repeating information. 

If a constant input length was used, slice duplication, or sub-sampling would be needed.
However,  duplication of slices, to increase the input length, is an ad-hoc procedure, since there does not exist any general pre-known  template for choosing which slices to duplicate. Moreover, sub-sampling  slices within a CT scan to reach a fixed number of slices is also ad-hoc, since significant  information for the final prediction could be discarded. 
We have examined these options in our experiments, achieving worse performance. We have not included these experiments in Table \ref{ablations}, so as to not clutter the presented results.

ii) Next we studied the effect of the   'alignment' step  on RACNet's performance. We compared the performance when the 'alignment' step was included, or not. Table \ref{ablations} illustrates that the 'alignment' case provided the best results. This result was expected, since it managed to better align the important slices in each CT scan series.

iii) Then we studied the effect of using various CNNs in RACNet. Table \ref{ablations} shows that, when EfficientNetB0 was used as the CNN part of RACNet, best performance was achieved, compared to the cases when other state-of-the-art CNNs were used, in particular ResNet-50 and DesneNet-121. We also studied the effect of using 3-D convolutions instead of a RNN in RACNet. Table \ref{ablations} shows that using the RNN provides a better performance, with the model being also lighter. In addition, we studied the effect of using different numbers of hidden units in the first dense layer of RACNet. Table \ref{ablations} illustrates that 128 units provided the best performance compared to the cases when 16 or 64 units were utilized.

Until this point, the presented results  refer to the case where no segmentation has been performed on the 3-D CT scan inputs); we did that as we did not want the specific selection of the lung segmentation method to affect the presented analysis and obtained results. In the following Subsection we examine the obtained performance of RACNet when its input are the segmented CT scans and compare its performance to the best performing methods in the above-mentioned ICCV Competition.

\begin{table}[t]
\caption{Performance comparison in various ablation studies on COV19-CT-DB database (non-segmented data)}
\label{ablations}
\centering
\scalebox{0.9}{
\begin{tabular}{|c|c|c|c|}
\hline
COV19-RACNet    & \multicolumn{2}{c|}{Accuracy} & \multicolumn{1}{c|}{F1 Score}\\
\hline
\hline
 & COVID & non COVID & \\ 
\hline
64 units in dense layer  & 0.79  & 0.83  &  0.86  \\
\hline
16 units in dense layer  & 0.78  & 0.82  &  0.85  \\
\hline
3D conv instead of RNN & 0.79  & 0.84  & 0.87  \\
\hline
ResNet-50 as CNN & 0.80  & 0.85  & 0.88  \\
\hline
DenseNet-121 as CNN & 0.79  & 0.84  &  0.87  \\
\hline
without 'alignment'  & 0.80  & 0.85  &  0.88  \\
\hline
without mask   & 0.78 & 0.84  &  0.87  \\
\hline
total (non-segmented data, 128 units in dense layer)   &\textbf{0.82} &  \textbf{0.86} &  \textbf{0.90}  \\
\hline
\end{tabular}
}
\end{table}

\subsection{Comparison with best performing methods in ICCV Competition }

As described in \cite{ref12}, the FDVTS-COVID network achieved the top performance in this Competition. It included a Periphery-aware Spatial Prediction network, which predicted whether a pixel belonged to the interior of the lung region, as well as the distance to the region boundary. This network was a pre-trained U-Net network with an encoder-decoder architecture; ResNet was adopted as the encoder. Each CT image was at first augmented and then fed into this encoder, generating vector representations. A classifier was trained on top of these representations for COVID-19 classification. Meanwhile, these representations were mapped by a projection network to new representations which were further enhanced in a contrastive learning manner.

The SenticLab.UAIC network ranked second, using an inflated 3D ResNet50 model with non local operations on the second and third layers. Inflated convolutions were obtained by expanding filters and pooling kernels of 2D ConvNets into 3D, resulting in learning spatio-temporal feature extractors from 3D images while using ImageNet architectures and label smoothing. To handle the variable length of CT-Scans, a sub-sampling technique, or padding, was used for lengths above, or under 128 respectively. During inference, parts of a single CT-Scan volume were inputted several times in the model; a threshold procedure followed to eliminate some results; final prediction was based on  majority voting over remaining results. 

The ACVLab network ranked third, based on either  slice-level, or 3D volume analysis. In the first case a vision Swin-Transformer was used for single-slice level classification followed by Wilcoxon signed-rank test. In the second case a Within-Slice-Transformer and a Between-Slice-Transformer were used, based on ResNet50 for feature extraction and self-attention for context-encoded features.

To compare the performance of RACNet with these methods, we first performed segmentation of  all 3-D CT scans. We used  a combination of morphological transforms  and a pre-trained U-Net model  \cite{ref14} resulting in a 2D semantic segmentation network. More specifically, for each CT-scan, every slice first passed through the pre-trained U-Net model. After all slices of the ct-scan were  segmented by the U-Net model, there was a checking procedure to assure that all slices were segmented. If a slice had a mask area less than 40 \% of the largest mask area of the CT-scan, then morphological transforms were used to segment this slice. Then RACNet was trained with the segmented data.

Table \ref{comparison_sota} compares the performance of these three networks and RACNet,  in terms of the criterion used in the Competition (Macro F1), whilst showing the F1 criterion values for both the COVID and non-COVID categories. It can be seen that RACNet outperforms all three networks, improving the total Macro F1 values, by an additional 3.5-5.0 \%. Moreover, it has a much larger increase in diagnosis of COVID-19 cases, which is very important, improving F1 by adding more than 10 \% to the best detection.

By comparing Tables \ref{3dcnn_rnn} and \ref{comparison_sota} one can see that the improvement in RACNet performance when using segmented CT scans compared to non-segmented ones is an additional 3.8 \% in F1 value. The accuracy values for COVID-19 and non-COVID categories, in the case of segmented inputs, were 0.88 and 0.89, compared to 0.82 and 0.86 in the non-segmented case respectively.

\begin{table}[t]
\caption{Competition Results on COV19-CT-DB (segmented data); F1 Score in \%  }
\label{comparison_sota}
\centering
\scalebox{1.}{
\begin{tabular}{ |c||c|c|c| }
 \hline
\multicolumn{1}{|c||}{\begin{tabular}{@{}c@{}} Networks \end{tabular}} & 
\multicolumn{1}{c|}{\begin{tabular}{@{}c@{}} Macro F1  \end{tabular}} &
\multicolumn{1}{c|}{\begin{tabular}{@{}c@{}}  F1 (COVID) \end{tabular}} &
\multicolumn{1}{c|}{\begin{tabular}{@{}c@{}} F1 (non-COVID) \end{tabular}} \\ 
  \hline
 \hline
FDVTS\_COVID &    \begin{tabular}{@{}c@{}}  90.43  \end{tabular} & \begin{tabular}{@{}c@{}} 83.60  \end{tabular} & \begin{tabular}{@{}c@{}} 97.27  \end{tabular} \\
\hline

SenticLab.UAIC &   \begin{tabular}{@{}c@{}}  90.06 \end{tabular} & \begin{tabular}{@{}c@{}}  82.96 \end{tabular} &
\begin{tabular}{@{}c@{}}  97.17    \end{tabular}  \\ 
\hline

ACVLab &   \begin{tabular}{@{}c@{}}  88.74    \end{tabular} & \begin{tabular}{@{}c@{}} 80.63    \end{tabular} &
\begin{tabular}{@{}c@{}} 96.84   \end{tabular} 
\\ 
\hline

RACNet &  
\begin{tabular}{@{}c@{}}   \textbf{93.83}    \end{tabular} 
& \begin{tabular}{@{}c@{}}   \textbf{93.62}    \end{tabular} 
& \begin{tabular}{@{}c@{}}   \textbf{94.04}   \end{tabular}
\\ 
\hline

\end{tabular}
}
\end{table}

\subsection{Comparison with 3-D CNNs on CC-CCII Database}

In the following, our aim is to test the effectiveness of RACNet on other databases. We consider the CC-CCII database \cite{zhang2020clinically}. The original CC-CCII dataset contains a total number of 617,775 slices of 6,752 CT scans obtained from 4,154 patients. However, there were some problems with it (i.e., damaged data, non-unified data type, repeated and noisy slices, disordered slices, and non-segmented slices). The authors of \cite{paperAAAI} published training and test partitions that did not include damaged data, naming this version of CC-CCII as 'Clean CC-CCII'. Clean CC-CCII is annotated in terms of COVID and non-COVID categories.

In order to handle the different number of input CT-scan length, the authors of \cite{paperAAAI} used two slice sampling algorithms: random sampling and symmetrical sampling. Specifically, the random sampling strategy was applied to the training set, which can be regarded as  data augmentation, while the symmetrical sampling strategy was performed on the test set to avoid introducing randomness into the testing results. The symmetrical sampling strategy referred to sampling from the middle to both sides at equal intervals. The relative order between slices remained the same before and after sampling. 

For a fair comparison we trained, fine-tuned and evaluated RACNet using the same Clean CC-CCII partitions. In particular: i) we trained the RACNet architecture with the Clean CC-CCII training set and tested its performance on the test set; ii) we pre-trained  RACNet  on COV19-CT-DB, then fine-tuned on the Clean CC-CCCII training set and tested its performance on the test set. 
Table \ref{3dcnn_rnn1} presents the performance of three state-of-art 3-D CNN networks, as well as of the above-described model, COVIDNet3D-L \cite{paperAAAI}. It also presents the performance of RACNet in the two above-mentioned contexts. 
The presented results in Table \ref{3dcnn_rnn1} indicate that RACNet greatly outperforms all three 3-D CNNs models, as well as the method proposed in \cite{paperAAAI}. It also indicates that the new COV19-CT-DB database can be used as an excellent prior for transfer learning and pre-training of deep neural networks for COVID-19 diagnosis in other medical environments.

In addition, we compared RACNet's performance to another network proposed in \cite{zhang2020clinically} which was trained using the original CC-CCII database. This network was trained for distinguishing NCP (COVID-19) from other common pneumonia and normal control CT scans. The utilized training set consisted of 752 covid patients, 797 common pneumonia patients and 697 normal controls. This model was evaluated on a test set consisting of 138 NCP, 135 common pneumonia and 129 normal control CT scans. The model achieved 92.49\% accuracy in the test set. However the authors of \cite{zhang2020clinically} did not publicize the exact data partitions. In order to obtain a fair comparison when comparing this model's performance to RACNet, we conducted a five-fold cross validation strategy, in each fold of which we kept the same numbers for the training and test sets as reported above. The accuracy obtained by RACNet for each fold was 96.1 \%, 96.9 \%, 96.2 \%, 95.6 \% and 95.9 \% respectively. The average accuracy of the five folds was 96.14 \%, outperforming by 3.6 \% the network proposed in \cite{zhang2020clinically}.  



\begin{table}[t]
\caption{Comparison of Performance of RACNet to that of 3-D CNNs on the CC-CCII Database (segmented data)}
\label{3dcnn_rnn1}
\centering
\scalebox{1.}{
\begin{tabular}{|c|c|}
\hline
Method   &  \multicolumn{1}{c|}{Accuracy score}\\
 \hline
\hline
Resnet3D101 \cite{paperAAAI}  &   89.62  \\
\hline

Densenet3D121 \cite{paperAAAI} & 88.97 \\
\hline

MCE\_18 \cite{paperAAAI} & 87.11 \\
\hline

COVIDNet3D-L \cite{paperAAAI} & 90.48 \\
\hline

RACNet & 93.64 \\
\hline

RACNet (pre-trained on COV19-CT-DB) & \textbf{95.33} \\
\hline
\end{tabular}
}
\end{table}

\subsection{Anchor Set Creation \&  Unification across Databases}

\begin{table}[h]
\caption{Description of the Severity Categories }
\label{tablecategories}
\centering
\scalebox{0.9}{
\begin{tabular}{| c | c | p{0.8\linewidth}|}

\hline
Category & Severity & Description\\
\hline
\hline
1 & Mild   & Few or no Ground glass opacities. Pulmonary parenchymal involvement $\leq 25 \%$ or absence \\
\hline
2 & Moderate  & Ground glass opacities.  Pulmonary parenchymal involvement  $25-50$\% \\
\hline
3 & Severe & Ground glass opacities. Pulmonary parenchymal involvement  $50-75$\% \\
\hline
4 & Critical &   Ground glass opacities. Pulmonary parenchymal involvement  $\geq 75$\%  \\
\hline
\end{tabular}
}
\end{table}

In the following we implemented the procedure of latent variable extraction and anchor set generation when training RACNet with 3206 CT scans (1634, i.e., almost all COVID-19 samples and a similar number, i.e., 1572 of non-COVID samples) from the COV19-CT-DB database. A validation set of 340 COVID-19 and 260 non-COVID-19 CT scans was used for selecting the best number of clusters.  As was already mentioned, this resulted in a set of 11 anchors, each represented by a vector in the 128-dimensional space. 7 of them corresponded to COVID-19 cases, with the rest corresponding to non COVID-19 cases. 

Table \ref{no_elems_0} provides the number of CT scans, belonging to RACNet training data, assigned to every generated cluster and their COVID-19, or non-COVID-19 category. It also provides a ranking of the severity of COVID-19, as classified by our medical experts, in the range from 1 to 4, with 4 denoting the critical status. Table \ref{tablecategories} describes each of these categories \cite{ref13}.  The centers of the above 11 clusters formed the anchor set generated during RACNet training on COV19-CT-DB database.

\begin{table}[t]
\caption{Number of elements per cluster, cluster category, Severity category}
\label{no_elems_0}
\centering
\scalebox{0.9}{
\begin{tabular}{|c|c|c|c|}
\hline
Cluster ID  & Number of CT Scans & Category  & Severity Category\\
\hline
\hline
0 & 231  & COVID-19 & 3\\
\hline
1 &  360  & COVID-19 & 2\\
\hline
2 & 344  & COVID-19 & 4\\
\hline
3 & 106 &  COVID-19 & 1\\ 
\hline
4 & 195 &  COVID-19 & 4\\ 
\hline
5 & 156 &  COVID-19 & 3\\  
\hline
6 & 242 &  COVID-19 & 4\\ 
\hline
7 & 107  &  non COVID-19 & 1\\ 
\hline
8 & 586 &   non COVID-19 & 1\\ 
\hline
9 & 557 & non COVID-19 & 1\\ 
\hline
10 & 322 & non COVID-19 & 1\\
\hline

\end{tabular}
}
\end{table}

For validation, we used this anchor set to classify the COV19-CT-DB test set. In particular, we fed each 3-D CT scan in the test set of the RACNet architecture; we extracted the  corresponding dense layer neuron outputs; we computed their euclidean distance from each anchor. Then they were classified according to the label of their nearest cluster center.
The obtained  classification performance over the test dataset was similar to the original RACNet's classification performance.


Moreover, our medical experts examined the 3-D scan inputs corresponding to the 11 cluster centres and produced justification for the respective diagnosis. 
Table \ref{no_elems_1} presents the findings detected in each cluster center.

\begin{table}[h]
\caption{Description of findings in each cluster center}
\label{no_elems_1}
\centering
\scalebox{0.88}{
\begin{tabular}{| c | p{0.95\linewidth}|}
\hline
Cluster ID  & Description\\
\hline
\hline
0 & Bilateral shadows ground-glass that become more compact locally in lower lung lobes  with an image of  pneumonia due to COVID-19; severe category  \\
\hline
1 &  Bilateral shadows ground-glass as in pneumonia due to COVID-19; moderate category  \\
\hline
2 & Minimal shadows ground-glass in left upper lung lobe. Severe thickening shadows,  dense atelectasis of lower lung lobes. Minimal pleural fluid on the right. Possible microbial cause; critical category  \\
\hline
3 &  Bilateral shadows ground-glass mainly in lower lung lobes  as in pneumonia due to COVID-19 in rather mild condition; mild category\\ 
\hline
4 & Bilateral shadows ground-glass that occupy more than 75 \% of the pulmonary parenchyma as in pneumonia COVID-19 of critical condition; critical category \\ 
\hline
5 & Bilateral shadows  ground-glass that occupy about 50 \%  of the pulmonary parenchyma as in pneumonia COVID-19 of critical condition; severe category \\  
\hline
6 & Bilateral shadows ground-glass that occupy more than 75 \% of the pulmonary parenchyma as in pneumonia COVID-19 of critical condition; critical category \\ 
\hline
7 & Bilateral emphysematous lesions as in chronic obstructive pulmonary disease. Dense atelectasis in paravertebral right lung; mild category \\ 
\hline
8 & Normal CT scan; mild category \\ 
\hline
9 & Normal CT scan; mild category \\ 
\hline
10 & Normal CT scan; mild category\\
\hline

\end{tabular}
}
\end{table}

\begin{figure*}[h!]
\centering
\includegraphics[width = 0.19\linewidth]{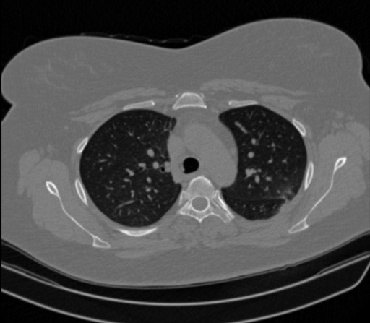}
\includegraphics[width = 0.19\linewidth,clip,trim={0 0.3cm 0 0}]{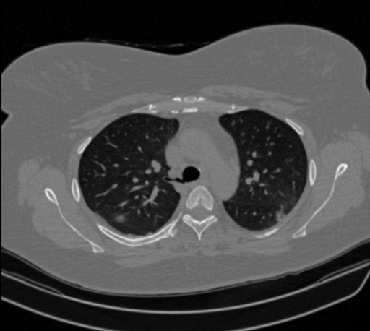}
\includegraphics[width = 0.19\linewidth]{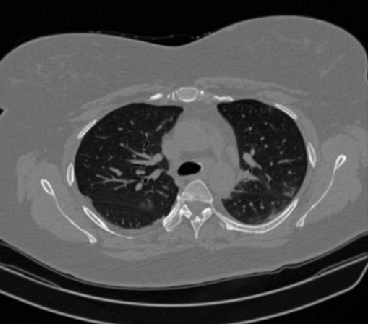}
\includegraphics[width = 0.19\linewidth]{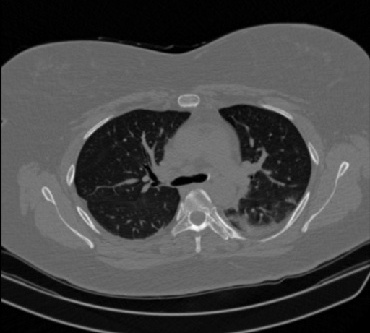}
\includegraphics[width = 0.19\linewidth]{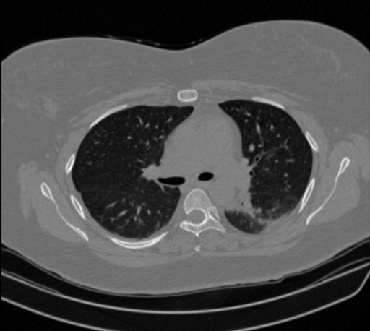} 
\\
\includegraphics[width = 0.19\linewidth]{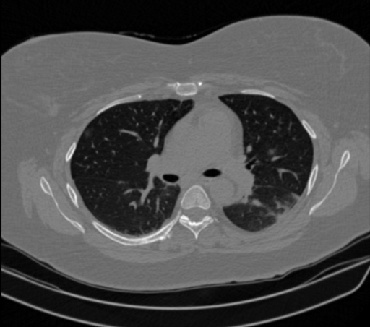}
\includegraphics[width = 0.19\linewidth,clip,trim={0 0.3cm 0 0}]{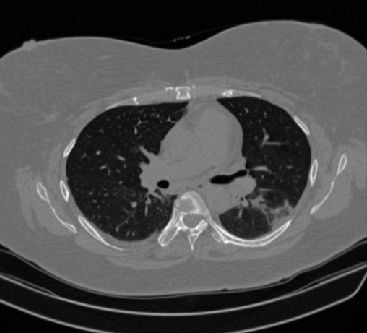}
\includegraphics[width = 0.19\linewidth]{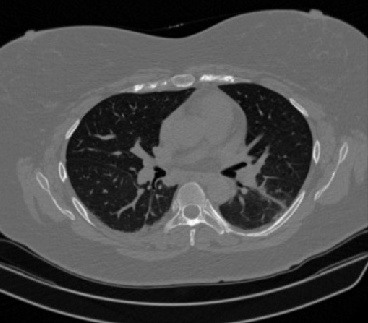}
\includegraphics[width = 0.19\linewidth]{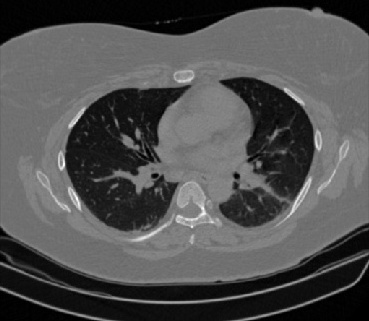}
\includegraphics[width = 0.19\linewidth]{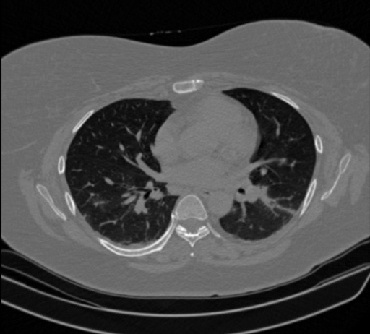}
\caption{Slices from COVID-19 cluster center 2, which is consistent with COVID-19 pneumonia bilateral thickening filtrates.  } 
\label{covid_0}
\end{figure*}

\begin{figure*}[h!]
\centering
\includegraphics[width = 0.19\linewidth]{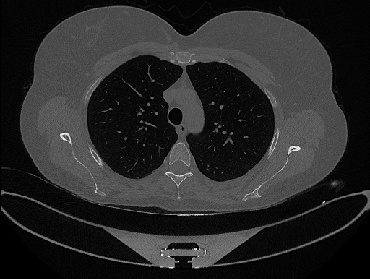}
\includegraphics[width = 0.19\linewidth]{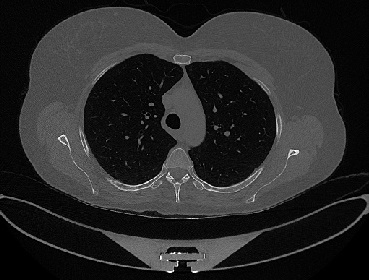}
\includegraphics[width = 0.19\linewidth]{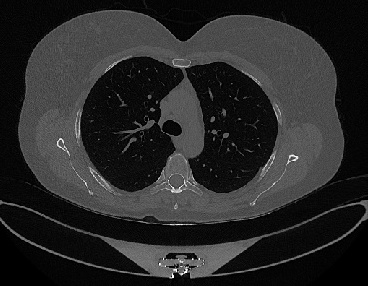}
\includegraphics[width = 0.19\linewidth]{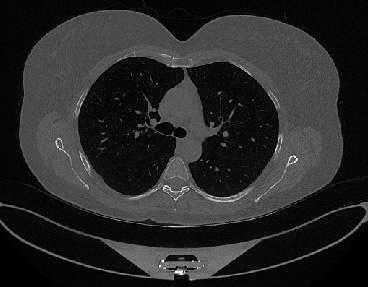}
\includegraphics[width = 0.19\linewidth]{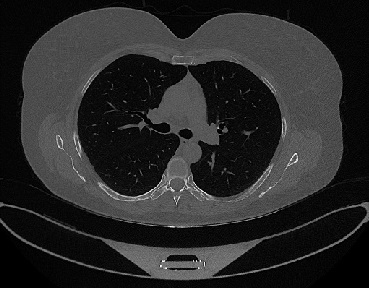}
\\
\includegraphics[width = 0.19\linewidth]{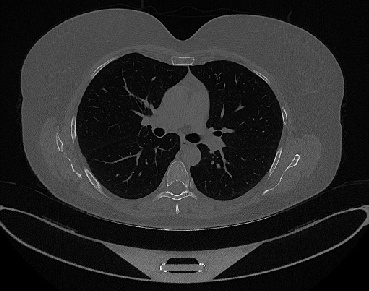}
\includegraphics[width = 0.19\linewidth]{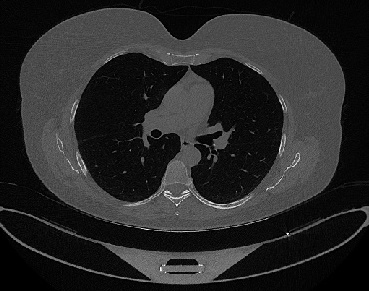}
\includegraphics[width = 0.19\linewidth]{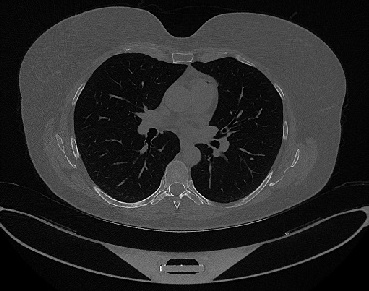}
\includegraphics[width = 0.19\linewidth]{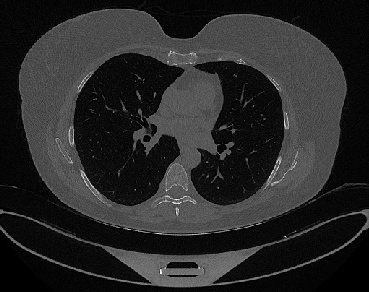}
\includegraphics[width = 0.19\linewidth]{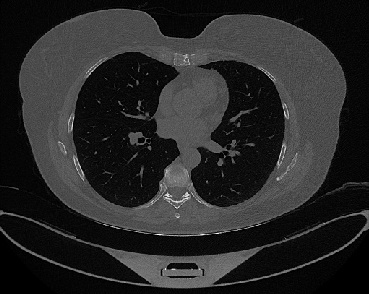}
\caption{Slices from  non COVID-19 cluster center 9.} 
\label{non_covid_1}
\end{figure*}

Some examples of CT slices from the cluster centers are given below. 
Figure \ref{clusters} shows 10 consecutive slices from  COVID-19 cluster center 0. Medical experts have annotated it as 'bilateral ground glass regions that appear, especially in lower lung lobes'. 
Figure \ref{covid_0} shows 10 slices from  COVID-19 cluster center 2. According to medical experts' annotation, this is consistent with 'COVID-19 pneumonia bilateral thickening filtrates'. 
Figure \ref{non_covid_1}, on the contrary, shows 10  slices from  non COVID-19 cluster center 9.

\begin{figure*}[h!]
\centering
\includegraphics[width = 0.19\linewidth]{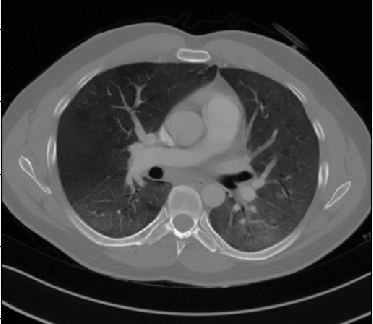}
\includegraphics[width = 0.19\linewidth]{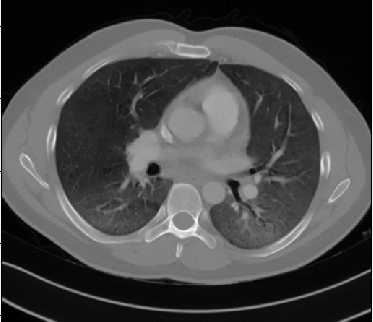}
\includegraphics[width = 0.19\linewidth]{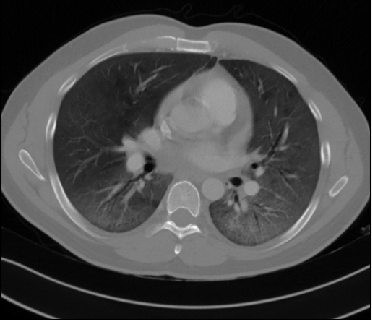}
\includegraphics[width = 0.19\linewidth]{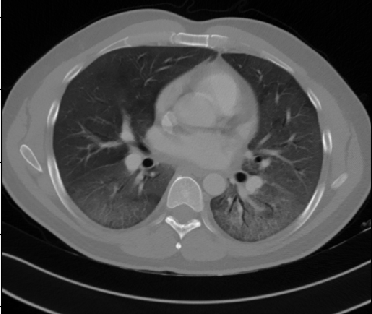}
\includegraphics[width = 0.19\linewidth]{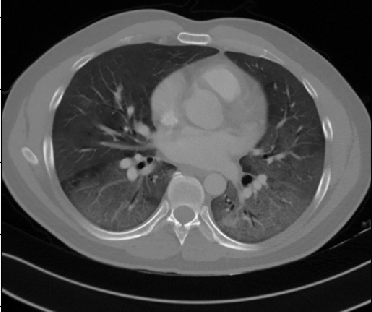} 
\\ 
\includegraphics[width = 0.19\linewidth,clip,trim={0 0.3cm 0 0}]{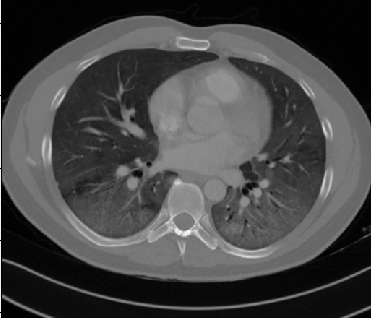}
\includegraphics[width = 0.19\linewidth]{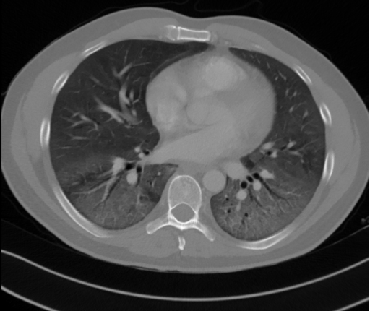}
\includegraphics[width = 0.19\linewidth]{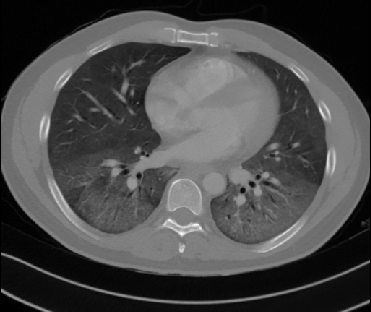}
\includegraphics[width = 0.19\linewidth]{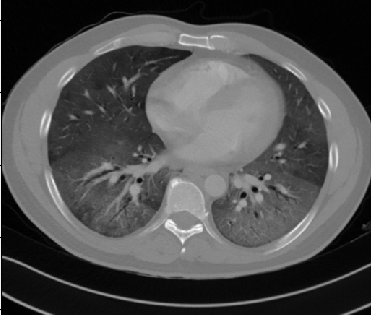}
\includegraphics[width = 0.19\linewidth]{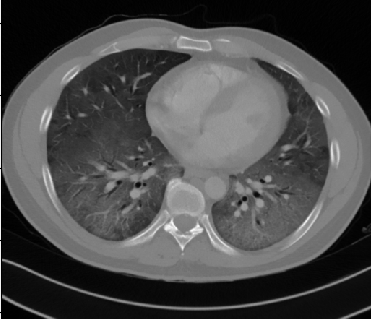}
\caption{Slices from  cluster  center 0 of COVID-19 category. Bilateral ground glass regions are seen especially in lower lung lobes.} 
\label{clusters}
\end{figure*}

The major advantage of the anchor set model is the insight that it introduces in the diagnosis process.
In each new test case, the generated decision is accompanied by the information about the anchor to which this case was assigned through the
above nearest neighbor classification procedure. As a result, the patient, or the doctor, can see which part of RACNet data-driven knowledge was used to make the specific diagnosis. 

In the following we used the RACNet trained on COV19-CT-DB and the set of 11 anchors for  unification with similar data-driven 
knowledge generated from another database, i.e., the aforementioned CC-CCII one.
We developed an efficient  unification procedure  based on the generated anchor set, which, on the one hand, alleviates the problem of 'catastrophic forgetting' when transfer learning is used and, on the other hand, reduces the high computational cost needed for retraining the deep learning model.

\begin{figure*}[h!]
\centering
\includegraphics[width = 0.15\linewidth]{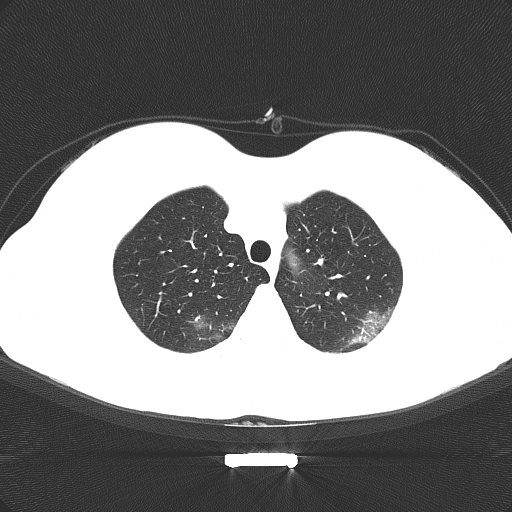}
\includegraphics[width = 0.15\linewidth]{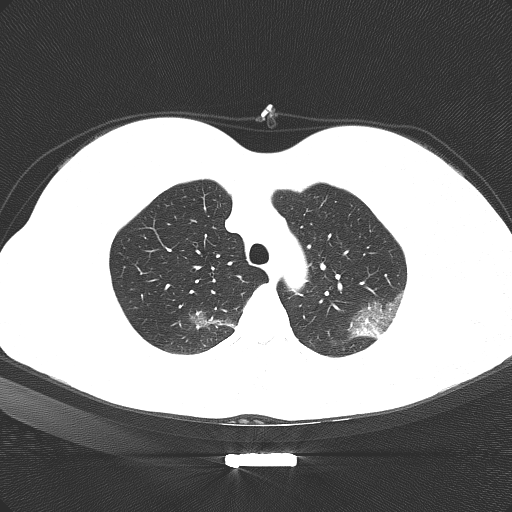}
\includegraphics[width = 0.15\linewidth]{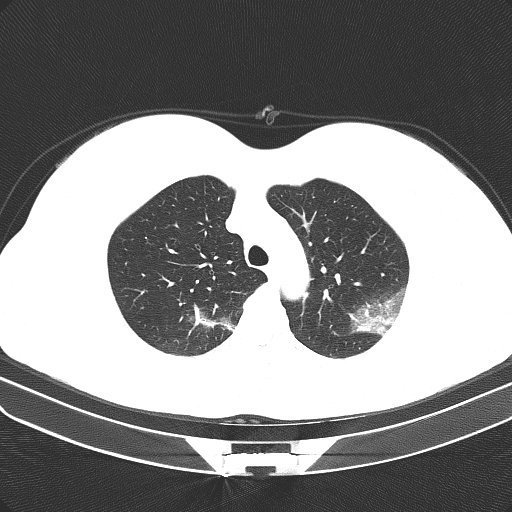}
\includegraphics[width = 0.15\linewidth]{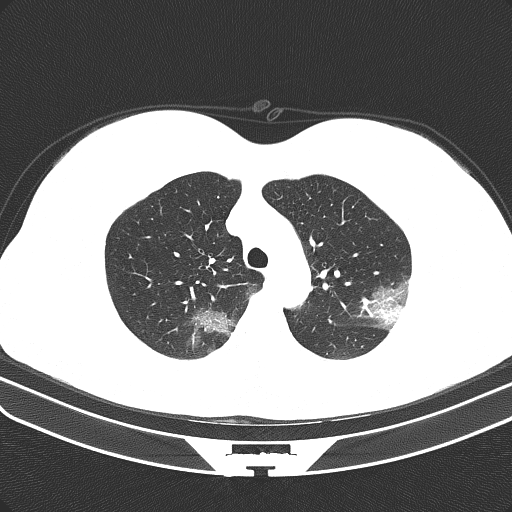}
\includegraphics[width = 0.15\linewidth]{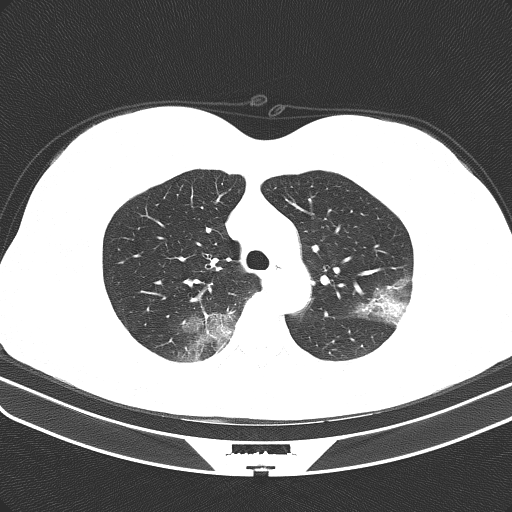} \\
\includegraphics[width = 0.15\linewidth]{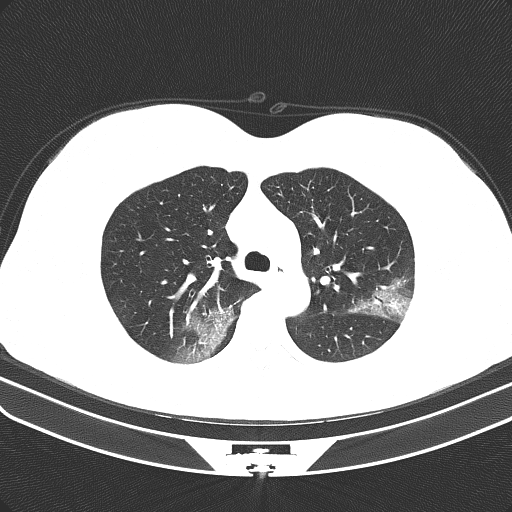}
\includegraphics[width = 0.15\linewidth]{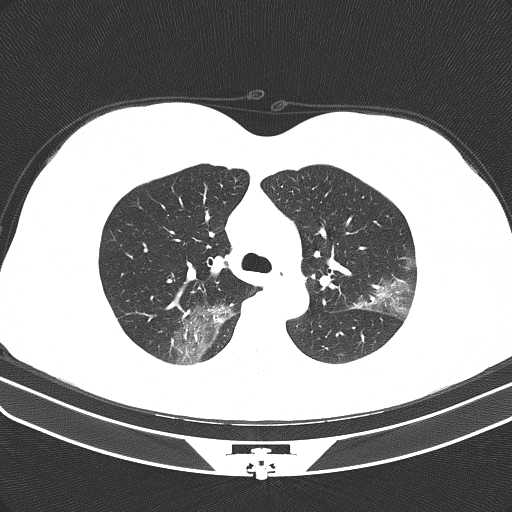}
\includegraphics[width = 0.15\linewidth]{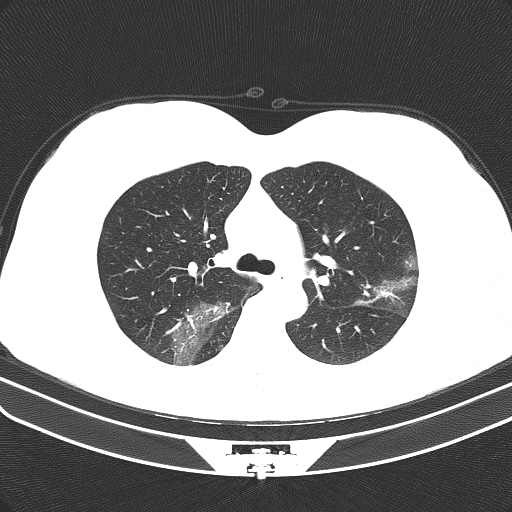}
\includegraphics[width = 0.15\linewidth]{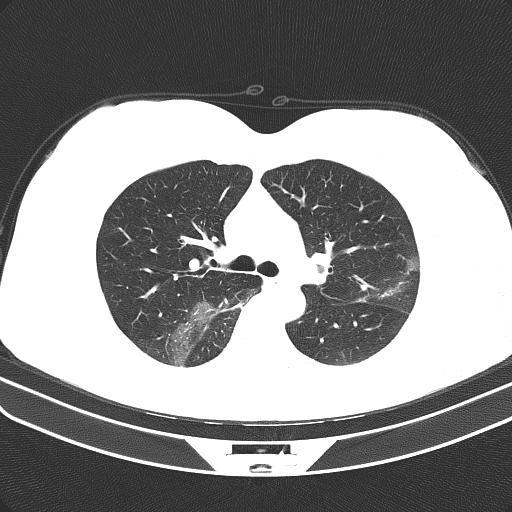}
\includegraphics[width = 0.15\linewidth]{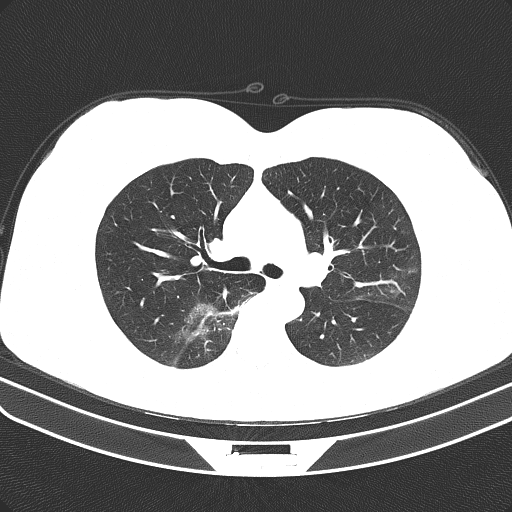}
\caption{Slices of a new COVID-19 anchor, with ground glass regions in the lungs.} 
\label{covid_russian}
\end{figure*}

Firstly, we computed the 128-dimensional features for each CT scan of the CC-CCII database using the RACNet's model which had been trained with the COV19-CT-DB training set. Then, these 128-dimensional features formed the input to train a neural network, say NN$^{(1)}$, consisting of 3 fully connected layers, so as to predict the Covid/non-COVID status of the CC-CCII data. The three layers included 64,  128 and 2 (output) neurons respectively.
In a similar way, as we extracted the 11 clusters from RACNet in \ref{clustersection}, we extracted a set of  representations from the layer with 128 neurons; then, through clustering, we generated another set of cluster centres. In this case, the number of cluster centers that produced the best performance over the CC-CCII test partition set was 13. Figure \ref{covid_russian}, shows 10  slices from one of the extracted COVID-19 cluster centers.

As a result, we have created: a) a set of 11 clusters \& respective cluster centers,  using COV19-CT-DB and RACNet, b) a set of 13 clusters \& respective cluster centers, using CC-CCII, RACNet and NN$^{(1)}$.

In the following, we generated a unified prediction model, by: a)  merging the 11 cluster centers from COV19-CT-DB and the 13 cluster centers from CC-CCII, b) using the RACNet -NN$^{(1)}$ as the combined test model. In particular, we could classify any CT scan in the COV19-CT-DB and CC-CCII databases, by passing it through RACNet -NN$^{(1)}$ and computing which one out of the 24 cluster centers was nearest to the extracted representation in the 128-dimensional space. The obtained performance was almost identical to the one obtained when processing each database independently. 
This result was achieved without exchanging any data between the holders of the two databases. It was only assumed that the RACNet and NN$^{(1)}$ networks and the cluster center representations in the 128-dimensional space were made available to each other. 

\section{Conclusions and Future work}

In this paper we have developed RACNet, a new deep neural architecture which: a) harmonizes analysis of 3-D image volumes consisting of different number of slices and annotated per volume, b) unifies decisions made over different input datasets, thus enriching data-driven knowledge and improving its trusted use. RACNet was successfully used for obtaining high performance in COVID-19 diagnosis based on chest 3-D CT scans, whilst permitting continual learning and avoiding catastrophic forgetting. 

Future work includes extension of the RACNet model and of the presented approach, so as to include uncertainty estimation and domain adaptation to a large variety of related applications, referenced in the Introduction of the paper.

\section*{Acknowledgment}

We would like to thank GRNET, National Infrastructures for Research \& Technology, for supporting us through  Project "Machaon -
Advanced networking and computational services to hospital units in public cloud environment".





\bibliography{mybibfile}

\end{document}